\documentclass[letterpaper,english,aps,prb,twocolumn,floatfix,showpacs,amsfonts,amssymb,superscriptaddress]{revtex4}

\usepackage[T1]{fontenc}
\usepackage[latin1]{inputenc}
\usepackage{graphics}
\usepackage{amssymb}
\usepackage{amsmath}
\usepackage{epsfig}

\newcommand{\beq}{\begin{equation}}
\newcommand{\eeq}{\end{equation}}
\newcommand{\bea}{\begin{eqnarray}}
\newcommand{\eea}{\end{eqnarray}}

\newcommand{\rmi}{{\rm i}}

\newcommand{\lasco}{La$_{2-x}$Sr$_{x}$CuO$_{4}$}

\begin{document}
\def\tende#1{\,\vtop{\ialign{##\crcr\rightarrowfill\crcr
\noalign{\kern-1pt\nointerlineskip} \hskip3.pt${\scriptstyle
#1}$\hskip3.pt\crcr}}\,}

\title{Charge stripe order from antiphase spin spirals in the spin-Fermion model}

\author{Marcello B. Silva Neto}

\affiliation{Institut f\"ur Theoretische Physik III, Universit\"at
Stuttgart, Pfaffenwaldring 57, 70550, Stuttgart, Germany.}

\begin{abstract}
We revisit the ground state of the spin-Fermion model within a
semiclassical approximation. We demonstrate that antiphase spin
spirals, or $\pi$-spirals, whose chirality alternates between
consecutive rows (or columns) of local moments, have, for
sufficiently high carrier concentration, lower energy than the
traditional Shraiman and Siggia spirals. Furthermore, $\pi$-spirals 
give rise to modulated hopping, anisotropic 1D transport, and 
charge density wave formation. Finally, we discuss the relevance 
of $\pi$-spirals to the physics of charge stripe formation in 
cuprates, such as {\lasco}.
\end{abstract}

\maketitle

\section{Introduction}

The determination of the ground state of a strongly correlated
system where conduction electrons interact with local magnetic
moments is an issue that is relevant to many different areas of
condensed matter physics. The usual starting point for such study
is the so called Kondo lattice model (KLM), where the local moments
are assumed to form a regular lattice. The KLM has become the
paradigm for the study of the physics of heavy Fermion (HF) compounds
and has been extensively investigated both theoretically
and with numerical methods (for recent references see, for example, 
Refs. \onlinecite{Sigrist} and \onlinecite{Assaad}). When generalized to
include a Heisenberg superexchange interaction, $J$, that favours
antiferromagnetic ordering among nearest neighbour local moments, 
the generalized KML, or yet, the spin-Fermion (SF) model, has been
argued to capture the essential physics of the high temperature
superconductors (HTSC).\cite{SF-Model}

Within the framework of the HTSC, the conduction electrons in the
SF model are provided by the holes doped into the $p_x$ and $p_y$
orbitals of the O$^{--}$ ions, while the local moments are
provided by the incomplete $3d_{x^2-y^2}$ shell of the Cu$^{++}$
ions. Although this microscopic picture is by now well
established, after 20 years of intense theoretical and numerical
research\cite{Kastner} the ground state of the HTSC is not yet
completely understood. It is believed that the true nature of the
ground state is determined by the competition between the
different energy scales of the model: the hopping integral, $t$,
the Kondo coupling, $J_K$, and the antiferromagnetic
superexchange, $J$, as well as by disorder. A large number of
possible ground states have been obtained in the literature,
including phase separation,\cite{Chubukov-Musaelian} charge
stripes formation,\cite{Zaanen-Gunnarsson} and spiral magnetic
order,\cite{SS,Kane-Read} among others.

From the experimental point of view, inelastic neutron scattering
experiments within the superconducting phase of {\lasco},
$x>5.5\%$, revealed that dynamical incommensurate (IC) spin
correlations coexist with superconductivity \cite{Yamada}. In
addition, it has also been observed experimentally the existence
of IC "charge" peaks accompanying the IC magnetic order, with
twice the incommensurability of the magnetic one.\cite{Tranquada}
These results were immediately and consistently interpreted as
evidence of the formation of charge stripes that act as antiphase
domain walls,\cite{Zaanen-Gunnarsson} and apparently rule out
other possible ground states such as Shraiman-Siggia
spirals,\cite{SS} which, despite being able to explain the IC
magnetic order, require a uniformly charged ground
state.\cite{Marcello}

In this brief report we revisit the ground state of the SF model
for cuprate superconductors within a semiclassical approximation.
We demonstrate that antiphase spin spirals, or $\pi$-spirals,
whose chirality alternates between consecutive rows (or columns)
of local moments, have, for sufficiently high carrier
concentration, lower energy than the traditional Shraiman and
Siggia spirals. Furthermore, $\pi$-spirals give rise not only to
IC magnetism but also to modulated hopping, anisotropic 1D
transport, and charge density wave (CDW) formation. Thus, our
theoretical results indicate, for the first time, the possibility 
of the {\it coexistence} between spiral magnetic order and CDW 
formation, in agreement with recent quantum Monte Carlo simulations 
for the SF model.\cite{Moreo} Finally, we discuss the relevance 
of $\pi$-spirals to the physics of charge stripe formation in 
the low temperature tetragonal phase of cuprate superconductors, 
such as {\lasco}.

\section{The Spin-Fermion Model}

The Hamiltonian for the spin-fermion model can be written as
$H=H_t+H_K+H_J$ where
\bea
H_t&=&-t\sum_{\langle
  i,j\rangle,\alpha}(p^\dag_{i,\alpha}p_{j,\alpha}+h.c.)
-\mu\sum_{i\alpha}p^\dag_{i,\alpha}p_{i,\alpha}
\nonumber\\
H_K&=&\frac{J_K}{2}\sum_{i,\alpha,\beta} {\bf S}_i \cdot
p^\dag_{i,\alpha}\,\sigma_{\alpha\beta}
\,p_{i,\beta}\nonumber\\
H_J&=&J\sum_{\langle i,j\rangle}{\bf S}_{i}\cdot{\bf S}_{j},
\label{Hamiltonian}
\eea
$p^\dag_{i,\alpha}$ creates a hole at the site $i$ with spin
projection $\alpha$, $t$ is the nearest-neighbour hopping
integral, $J_K$ is an antiferromagnetic exchange interaction
between the Cu$^{++}$ local moment and the spin of the doped
O$^{--}$ holes, $J$ is the Heisenberg antiferromagnetic
superexchange, and $\mu$ is the chemical potential. In the large
$J_K/t$ limit, the above Hamiltonian is known to reduce to the
$t-J$ model.

Since conduction electrons and local moments in the SF model are
independent degrees of freedom we can treat the local moments
classically without affecting the mobile Fermions.\cite{Hamada}
This is the basis of the semiclassical approximation used here. In
this case, the Hamiltonian becomes quadratic in the Fermion fields
and can be diagonalized. For simplicity we choose a
spin-quantization basis such that ${\bf S}_i=(0,0,S)$ at every
site, and the Kondo term reduces to a shift on the chemical
potential for the different spin components of the doped holes,
$H_K=(J_K/2) S\sum_{i}(p^\dag_{i,\uparrow}p_{i,\uparrow}-
p^\dag_{i,\downarrow}p_{i,\downarrow})$.

As it was originally proposed long ago by Shraiman and
Siggia,\cite{SS} for $t>J$ (for {\lasco} $t/J\approx 3$) the
hopping of the doped holes is favoured by a noncollinear
configuration for the local moments of the spiral type (see also
Ref.\ \onlinecite{SK}). Because of our choice for the local
spin-quantization basis, as the spins spiral so does the local
frame. This will affect the mobile holes that transform under the
following SU$(2)$ transformation
\beq U({\bf x}_i)=e^{\rmi\theta_i {\bf \zeta}\cdot\vec{\sigma}/2}=
\cos{\frac{\theta_i}{2}}+\rmi\;
\vec{\sigma}\cdot{\bf
\zeta}\sin{\frac{\theta_i}{2}}. \eeq
Here ${\bf \zeta}$ is a unitary vector pointing to an arbitrary
direction in the plane perpendicular to the local spin, and
$\vec{\sigma}$ are the Pauli matrices. Under such SU$(2)$
transformation the Fermion fields are then transformed as
\beq
\left( \begin{array}{c}
     p_{i,\uparrow} \\ p_{i,\downarrow}
           \end{array} \right)=U^\dag({\bf x}_i)
\left( \begin{array}{c}
     c_{i,\uparrow} \\ c_{i,\downarrow}
           \end{array} \right),
\eeq
and we usually write
\beq
\theta_i={\bf q}\cdot{\bf x}_i,
\eeq
in such a way that the classical AF N\'eel state is parametrized
by ${\bf q}=(\pi,\pi)$ and has classical energy $E_{AF}=-4NJS^2$,
where $N$ is the total number of local moments.

\section{Inphase or $0$-spirals}

\begin{figure}[t]
\begin{center}
\includegraphics[scale=0.45]{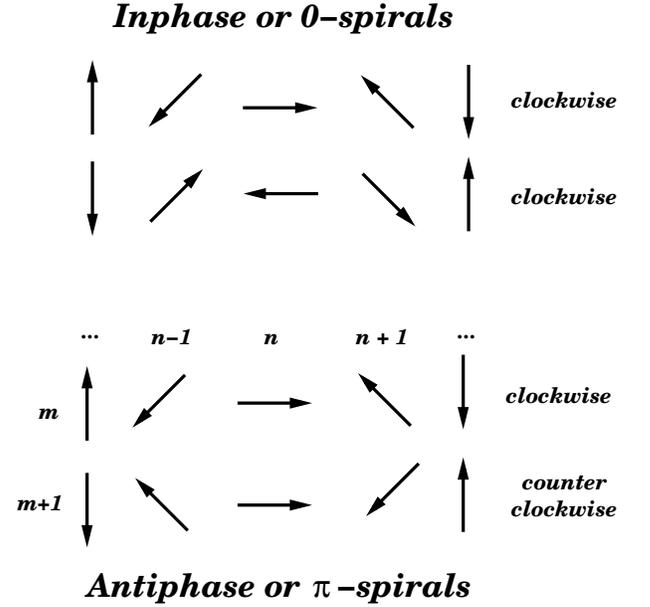}
\caption{Inphase (top) and antiphase (bottom) spiral
configurations for
  the magnetic ground state of the spin-Fermion model. For inphase
  $(1,0)$ spirals all spins rotate clockwise, for example, in every
  row $m$ of copper ions. For antiphase $(1,0)$ spirals, on contrary,
  the sense of rotation alternates between two consecutive rows,
  $m$ and $m+1$ for example.}
\label{Fig-Spiraling}
\end{center}
\end{figure}

Inphase or $0$-spirals are parametrized as having an IC wave vector
given by, for example,
\beq
{\bf q}=(q_x,\pi).
\eeq
This corresponds to Shraiman-Siggia original solution\cite{SS}
(see Fig.\ \ref{Fig-Spiraling}) and in this case the SF model
reduces to
{\small
\bea
H_t&=&-t\sum_{\langle
  i,j\rangle,\alpha}
\cos{\left(\frac{{\bf q}\cdot({\bf x}_i-{\bf x}_j)}{2}\right)}
(c^\dag_{i,\alpha}c_{j,\alpha}+h.c.)\nonumber\\
&-&t\sum_{\langle i,j\rangle}
\sin{\left(\frac{{\bf q}\cdot({\bf x}_i-{\bf x}_j)}{2}\right)}
(\rmi e^{-\rmi\zeta}c^\dag_{i,\uparrow}c_{j,\downarrow}-
 \rmi e^{\rmi\zeta}c^\dag_{j,\uparrow}c_{i,\downarrow}),\nonumber\\
H_K&=&\frac{J_K S}{2}\sum_{i} (c^\dag_{i,\uparrow} c_{i,\uparrow}-
c^\dag_{i,\downarrow} c_{i,\downarrow}),\nonumber\\
H_J&=&-2NJS^2\left[1-\cos{q_x}\right].
\label{0-Hamiltonian}
\eea
}
The spiraling of the local moments favours the hopping of the
conduction electrons in the direction of the pitch of the spiral
via the $H_t$ term, at the price of some magnetic energy loss
($H_J$ in Eq. (\ref{0-Hamiltonian}) is larger than $-4NJS^2$). In
momentum space the above Hamiltonian reads
{\small
\beq
H= \sum_{\bf k} ( \begin{array}{cc}
     c^\dag_{{\bf k},\uparrow} & c^\dag_{{\bf k},\downarrow}
           \end{array} )
\left( \begin{array}{cc}
     \xi_0({\bf k})+\frac{J_K S}{2} & \xi_2({\bf k}) \\
     \xi_2({\bf k}) & \xi_0({\bf k})-\frac{J_K S}{2} \\
              \end{array} \right)
\left( \begin{array}{c}
     c_{{\bf k},\uparrow} \\ c_{{\bf k},\downarrow}
           \end{array} \right),
\eeq
}
where
\bea
\xi_0({\bf k})&=&-2 t \cos{k_x}\cos{\frac{q_x}{2}},\nonumber\\
\xi_2({\bf k})&=&-2 t \left[\sin{k_x}\sin{\frac{q_x}{2}}+\sin{k_y}\right].
\eea
The above Hamiltonian can be diagonalized in momentum space and we
obtain the dispersions
\beq \epsilon^{\pm}_0({\bf k})=\xi_0({\bf k})
\pm\sqrt{\xi_2^2({\bf k})+\left(\frac{J_K S}{2}\right)^2}, 
\label{Bands} 
\eeq
with energy minima are located near ${\bf k}_0=(\pi/2,\pi/2)$ and
symmetry related points in the magnetic Brillouin zone.

%
%

We now calculate the total energy of the $0$-spiral ground state.
It is clear that, although the system pays magnetic energy in
order to stabilize the $0$-spiral, the gain in kinetic energy
\beq
E^0_{kin}=\sum_{{\bf k}}\epsilon^{-}_0({\bf k})\Theta(\mu-\epsilon^{-}_0({\bf k})),
\eeq
is {\it always larger} than the magnetic energy loss, already for
infinitesimal doping (as it has been demonstrated in Ref.\
\onlinecite{Juricic}, a Dzyaloshinskii-Moriya term, not considered
here, can shift this instability to finite doping, thus
stabilizing the N\'eel ground state). Here we assumed that only
the lower energy band in Eq.\ (\ref{Bands}) is filled, due to the
large gap $(J_K S)/2$ (recall that $J_K/t\gg 1$), and we used that
$k_F=\sqrt{2\pi \delta}$, where $\delta$ gives the carrier
concentration.

The are two problems with the $0$-spiral state: i) it does not
break translational invariance in the charge sector, as it is
observed experimentally;\cite{Tranquada} ii) it favours the
transport in the direction {\it parallel} to the pitch of the
spiral, while experiments have demonstrated that the anisotropic
1D transport should be {\it perpendicular} to it.\cite{Liu} As we
shall demonstrate now, both issues are naturally incorporated by
the antiphase or $\pi$-spiral state.

%
\begin{figure}[t]
\begin{center}
\includegraphics[scale=0.325]{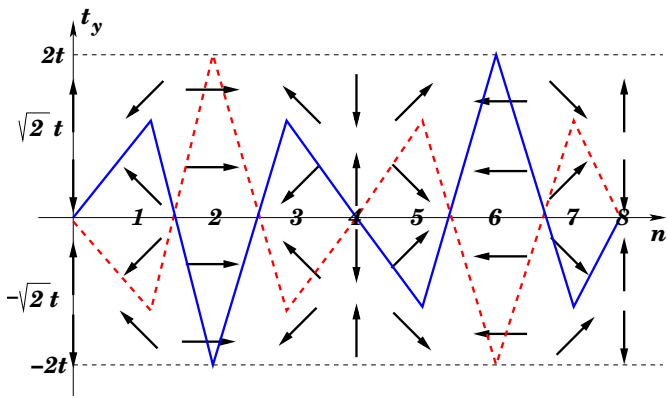}
\caption{(Color online) Modulated hopping amplitudes for the two
degenerate antiphase or $\pi$-spiral configurations at filling
$\delta=1/8$. The blue (solid) curve is the hopping amplitude when
the spins in the top row rotate clockwise (case drawn) while the
red (dashed) curve is the hopping amplitude when the spins in the
top row rotate counter-clockwise (not shown).}
\label{Fig-Modulated-Hopping}
\end{center}
\end{figure}
%

\section{Antiphase or $\pi$-spirals }

Antiphase or $\pi$-spirals can be parametrized by the IC wave
vector
\beq
{\bf q}_m=((-)^m q_x,\pi),
\eeq
where $m$ labels a certain row of Cu$^{++}$ ions, see Fig.\
\ref{Fig-Spiraling}. The spin-Fermion model in this case is
rewritten as ($m$ and $n$ label, respectively, rows and columns of
local moments)
\bea
H_{t}&=&-t\sum_{m,n,\alpha}
\cos{\frac{q_x}{2}}
(c^\dag_{(m,n+1)\alpha}c_{(m,n)\alpha}+h.c.)\nonumber\\
&-&t\sum_{m,n,\alpha}
\sin{(n q_x)}
(c^\dag_{(m+1,n)\alpha}c_{(m,n)\alpha}+h. c.)\nonumber\\
&+&t\sum_{m,n}
\cos{(n q_x)}
(\rmi e^{-\rmi\zeta}c^\dag_{(m+1,n)\uparrow}c_{(m,n)\downarrow}+h. c.)
\nonumber\\
&+&t\sum_{m,n}
\cos{(n q_x)}
(\rmi e^{\rmi\zeta}c^\dag_{(m+1,n)\downarrow}c_{(m,n)\uparrow}+h. c.)\nonumber\\
H_K&=&\frac{J_K S}{2}\sum_{m,n} (c^\dag_{(m,n)\uparrow} c_{(m,n)\uparrow}-
c^\dag_{(m,n)\downarrow} c_{(m,n)\downarrow})\nonumber\\
H_J&=&2 N J S^2 \cos{q_x}.
\label{Pi-Hamiltonian}
\eea

There are important differences with respect to the case of the
$0$-spirals: i) in the $\pi$-spiral ground state, the spin
independent part of the hopping {\it perpendicular} to the spiral
pitch becomes {\it modulated}, see Fig.\ \ref{Fig-Modulated-Hopping}; 
ii) it becomes maximal for parallel configurations of the local 
moments, as in positions $n=2,6$; iii) it vanishes for antiparallel 
configurations for the local moments, as in positions $n=0,4,8$. 
Furthermore, the modulation in the vertical hopping leads to a 
modulation on the carrier density, and, as a consequence, to a 
CDW instability of the stripe sort.\cite{Zaanen-Gunnarsson} Finally, 
it is clear that the vertical hopping can be much larger than the 
horizontal one, thus exposing the 1D nature of the transport in 
the $\pi$-spiral phase.

We still have to show that the $\pi$-spiral phase has lower energy
than the $0$-spiral one. In momentum space the Hamiltonian reads
\begin{widetext}
\beq H= \sum_{\bf k} ( \begin{array}{cccc}
     c^\dag_{{\bf k},\uparrow} & c^\dag_{{\bf k+q},\uparrow} & c^\dag_{{\bf k},\downarrow} & c^\dag_{{\bf k+q},\downarrow}
           \end{array} )
\left( \begin{array}{cccc}
\xi_0({\bf k})+\frac{J_K S}{2} & -\Delta({\bf k}) & e^{-\rmi\zeta}\xi_2({\bf k}) & e^{-\rmi\zeta} \Delta({\bf k})  \\
-\Delta({\bf k}) & \xi_0({\bf k+q})+\frac{J_K S}{2} & e^{-\rmi\zeta} \Delta({\bf k}) & e^{-\rmi\zeta}\xi_2({\bf k+q}) \\
e^{\rmi\zeta}\xi_2({\bf k}) & e^{\rmi\zeta} \Delta({\bf k}) & \xi_0({\bf k})-\frac{J_K S}{2} & -\Delta({\bf k}) \\
e^{\rmi\zeta} \Delta({\bf k}) & e^{\rmi\zeta}\xi_2({\bf k+q}) & -\Delta({\bf k}) & \xi_0({\bf k+q})-\frac{J_K S}{2}
\end{array} \right)
\left( \begin{array}{c}
     c_{{\bf k},\uparrow} \\ c_{{\bf k+q},\uparrow} \\ c_{{\bf k},\downarrow} \\ c_{{\bf k+q},\downarrow}
           \end{array} \right).
\eeq
\end{widetext}
%

%
\begin{figure}[t]
\begin{center}
\includegraphics[angle=-90,scale=0.32]{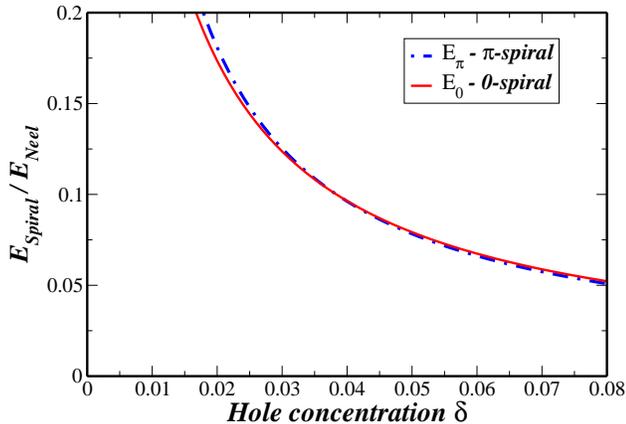}
\caption{Doping dependence of the energy of both inphase or
  $0$-spirals and antiphase or $\pi$-spirals. We see that
  although $E_{\pi}>E_0$ at low doping, for larger $\delta$ it 
  becomes the true semiclassical magnetic ground state of the system.}
\label{Fig-Energy-Spiral}
\end{center}
\end{figure}
%

The above Hamiltonian can be diagonalized and the new
dispersions are
\beq
\epsilon^{(\pm,\pm)}_\pi({\bf k})=\pm\sqrt{F_I({\bf k})\pm2\sqrt{F_{II}({\bf k})}},
\label{New-Bands}
\eeq
where
\beq
F_I({\bf k})=\xi_2^2({\bf k})+\xi_0^2({\bf k})+
2\Delta^2({\bf k})+\left(\frac{J_K S}{2}\right)^2,
\eeq
and
\beq
F_{II}({\bf k})=(\Delta^2({\bf k})-\xi_2({\bf k})\xi_0({\bf k}))^2+
\left(\frac{J_K S}{2}\right)^2(\xi_0^2({\bf k})+\Delta^2({\bf k})).
\eeq
Here
\beq
\Delta({\bf k}) = t \sin{k_y}
\eeq
is the CDW gap, with the property $\Delta({\bf k+q})=-\Delta({\bf k})$. We see
that Eqs. (\ref{New-Bands}) reduce to Eqs.\ (\ref{Bands}) in the
limit $\Delta=0$. For $\Delta\neq 0$, in turn, both
$\epsilon^{(-+)}_\pi({\bf k})$ and $\epsilon^{(--)}_\pi({\bf k})$
are always smaller than $\epsilon^{-}_0({\bf k})$, and thus
\bea
E^\pi_{kin}&=&\frac{1}{2}\sum_{{\bf k}}\left\{\epsilon^{(-+)}_\pi({\bf k})
\Theta(\mu-\epsilon^{(-+)}_\pi({\bf k}))\right.\nonumber\\
&+&\left. \sum_{{\bf k}}\epsilon^{(--)}_\pi({\bf k})
\Theta(\mu-\epsilon^{(--)}_\pi({\bf k}))\right\},
\eea
provides us with a rather large gain in kinetic energy. In particular, 
we found that for $\delta \approx 5\%$, $\pi$-spirals already have 
lower energy than $0$-spirals, see Fig.\ \ref{Fig-Energy-Spiral}. 
We used units such that $J=1$, with $t=3$ and $J_K=5$, which have 
the correct hierarchy tipically observed in superconducting cuprates 
($J_K>t>J$). As a result, 
the stabilization of a striped CDW phase coexisting with $\pi$-spirals 
breaks the translational symmetry in the charge sector, gives rise 
to IC magnetic correlations, and favours 1D trasport {\it perpendicular} 
to the spiral pitch.

\section{Conclusions}

Using a semiclassical approximation we have revisited the ground
state of the SF model. We have shown that $\pi$-spirals have lower
energy than $0$-spirals and give rise to modulated hopping,
anisotropic transport, and CDW formation. Thus, although such
semiclassical analysis is, strictly speaking, only valid for large
$S$, the possibility of the coexistence of spiral magnetic order
and charge modulation is an appealing feature of the new
semiclassical ground state here presented, captures the essential
physics of the charge and magnetic IC order in {\lasco}, and has
been recently obtained numerically with quantum Monte
Carlo.\cite{Moreo} The stability of the semiclassical $\pi$-spiral
ground state towards fluctuations has still to be demonstrated,
but we believe that, as it happens for the case of $0$-spirals in
the $t-J$ model,\cite{SK} the $\pi$-spiral phase can also be made
robust through the inclusion of next-to-nearest neighbour hopping
terms.

The author acknowledges discussions with R.~Doretto, C.~Morais~Smith, 
A.~Moreo, O.~Sushkov, and J.~Zaanen.


\end{document}